\documentclass[twocolumn,superscriptaddress,footinbib,amsmath,amssymb,aps,prb,letterpaper,10pt,longbibliography,floatfix]{revtex4-2}

\usepackage[utf8]{inputenc}

\usepackage{amsfonts}
\usepackage{braket}
\usepackage{dsfont}
\usepackage{graphicx}
\usepackage{xcolor}
\usepackage{listings}
\usepackage{comment}
\usepackage{hyperref}
\usepackage{url}

\usepackage[caption=false]{subfig}

\begin{document}

\title{Measuring coherence factors of states in superconductors through local current}%
    \author{Rodrigo A. Dourado}
\affiliation{ \textit{Instituto de F\'isica de S\~ao Carlos, Universidade de S\~ao Paulo, 13560-970 S\~ao Carlos, S\~ao Paulo, Brazil}	}
\affiliation{ \textit{Departamento de Física, Universidade Federal de Minas Gerais, C. P. 702, 30123-970, Belo Horizonte, MG, Brazil}	}

\author{Jeroen Danon}
\affiliation{Department of Physics, Norwegian University of Science and Technology, Trondheim NO-7491, Norway}

\author{Martin Leijnse}
\affiliation{Division of Solid State Physics and NanoLund, Lund University, S-22100 Lund, Sweden}

\author{Rub\'{e}n Seoane Souto}
\affiliation{Instituto de Ciencia de Materiales de Madrid (ICMM), Consejo Superior de Investigaciones Cient\'{i}ficas (CSIC), Sor Juana In\'{e}s de la Cruz 3, 28049 Madrid, Spain}

\begin{abstract}
    The coherence factors of quasiparticles in a superconductor determine their properties, including transport and susceptibility to electric fields. In this work, we propose a way to infer the local coherence factors using local transport to normal leads. Our method is based on measuring the local current through a lead as the coupling to a second one is varied: the shape of the current is determined by the ratio between the local coherence factors, becoming independent of the coupling to the second lead in the presence of local electron-hole symmetry, {\it i.e.} coherence factors $|u|=|v|$. We apply our method to minimal Kitaev chains: arrays of quantum dots coupled via narrow superconducting segments. These chains feature Majorana-like quasiparticles (zero-energy states with $|u|=|v|$) at discrete points in parameter space. We demonstrate that the local current allows us to estimate the local Majorana polarization (MP)--a measurement of the local Majorana properties of the state. We derive an analytical expression for the MP in terms of local currents and benchmark it against numerical calculations for 2- and 3-sites chains that include a finite Zeeman field and electron-electron interactions. These results provide a way to quantitatively assess the quality of Majorana states in short Kitaev chains.
\end{abstract}
	
\maketitle

\section{Introduction}

The electron-hole superposition of quasiparticles lies at the core of superconductivity, giving rise to key phenomena such as Andreev reflection and supercurrent interference~\cite{tinkham1975introduction}. The electron and hole character of a quasiparticle is described by the spatially dependent coherence factors 
$u$ and $v$. These coefficients determine the quasiparticles’ response to electric and magnetic fields, making them central to the behavior of superconducting systems and their applications in quantum technologies.

A Majorana bound state (MBS) is a special kind of subgap state that appears in topological superconductors and features an equal superpositions of electrons and holes, fulfilling $|u|=|v|$, and lies at zero energy~\cite{leijnse2012introduction, alicea2012new, lahtinen2017short, aguado2017majorana, beenakker2020search, prada2020andreev, flensberg2021engineered, das2023search, Souto_chapter,kouwenhoven2024perspective}. The electron-hole symmetry means that the quantum information encoded in a MBS cannot be read out with local probes, being, therefore, robust against local perturbations. MBSs are predicted to emerge as zero-energy excitations in p-wave superconductors \cite{Kitaev2001}. In the last decades, there has been an intense search for MBSs in semiconductor-superconductor systems~\cite{lutchyn2010majorana, oreg2010helical}, driven by their potential applications for topological quantum computation~\cite{kitaev2003fault, nayak2008non}. However, it has been proven to distinguish topological from trivial states due to, {\it e.g.} disorder or smooth confinement potentials~\cite{kells2012near, liu2017andreev, pan2021quantized, hess2021local, das2021disorder, das2023spectral}. 

Artificial Kitaev chains, composed of quantum dots (QDs) coupled to superconductors, have emerged as promising alternatives to realize MBSs~\cite{Sau2012,Leijnse2012, fulga2013adaptive,Miles2023,Samuelson_PRB2024}. In its minimal form, two QDs connect via a narrow superconducting segment that features spin-orbit coupling~\cite{bordin2023crossed, Dvir2023, wang2023triplet, zatelli2024robust, bordin2024signatures, ten2024two, haaf2024edgebulkstatesthreesite,Loo_arXiv2025}. The superconductor mediates the coupling between the QDs, allowing for crossed Andreev reflections (CAR) of Cooper pairs and elastic cotunneling (ECT). An external magnetic field is used to polarize the QDs. Using these ingredients, the low-energy properties of the system map to an effective 2-site Kitaev chain~\cite{liu2022tunable,tsintzis2022creating}. The system features well-localized MBSs at discrete points in parameter space (sweet spots). Although these states can be used to study the coherent properties of MBSs, including braiding and fusion protocols~\cite{liu2023fusion, Tsintzis2024, pandey2024nontrivial,Pino_PRB2024,Pan_PRB2025,Vimal_arXiv2025}, they are not topologically protected. For this reason, they are known as ``poor man's Majorana bound states"~\cite{Leijnse2012}. Topological protection can be achieved by increasing the number of sites in the Kitaev chains~\cite{Sau2012, bordin2024signatures, svensson2024quantum, haaf2024edgebulkstatesthreesite, dourado2025twositekitaevsweetspots}.

While many experiments reported the onset of zero-energy states in artificial Kitaev chains~\cite{bordin2023crossed, Dvir2023, wang2023triplet, zatelli2024robust, bordin2024signatures, ten2024two, haaf2024edgebulkstatesthreesite}, the assessment of the separation between the MBSs has been less explored. The local and nonlocal conductance have been used to characterize sweet spots~\cite{Dvir2023,Haaf2023}. Alternatively, the conductance quantization~\cite{Alvarado_PRB2024} and an additional QD provides information about the MBSs coupling at the end of the system~\cite{Deng_Science2016,Prada_PRB2017,Clarke_PRB2017,Deng_PRB2018,Seoane2023,Bordin_arXiv2025}. Although these proposals provide a qualitative understanding of the MBS localization, its quantitative measurement is still lacking.

\begin{figure}
\centering
\includegraphics[width=1\linewidth]{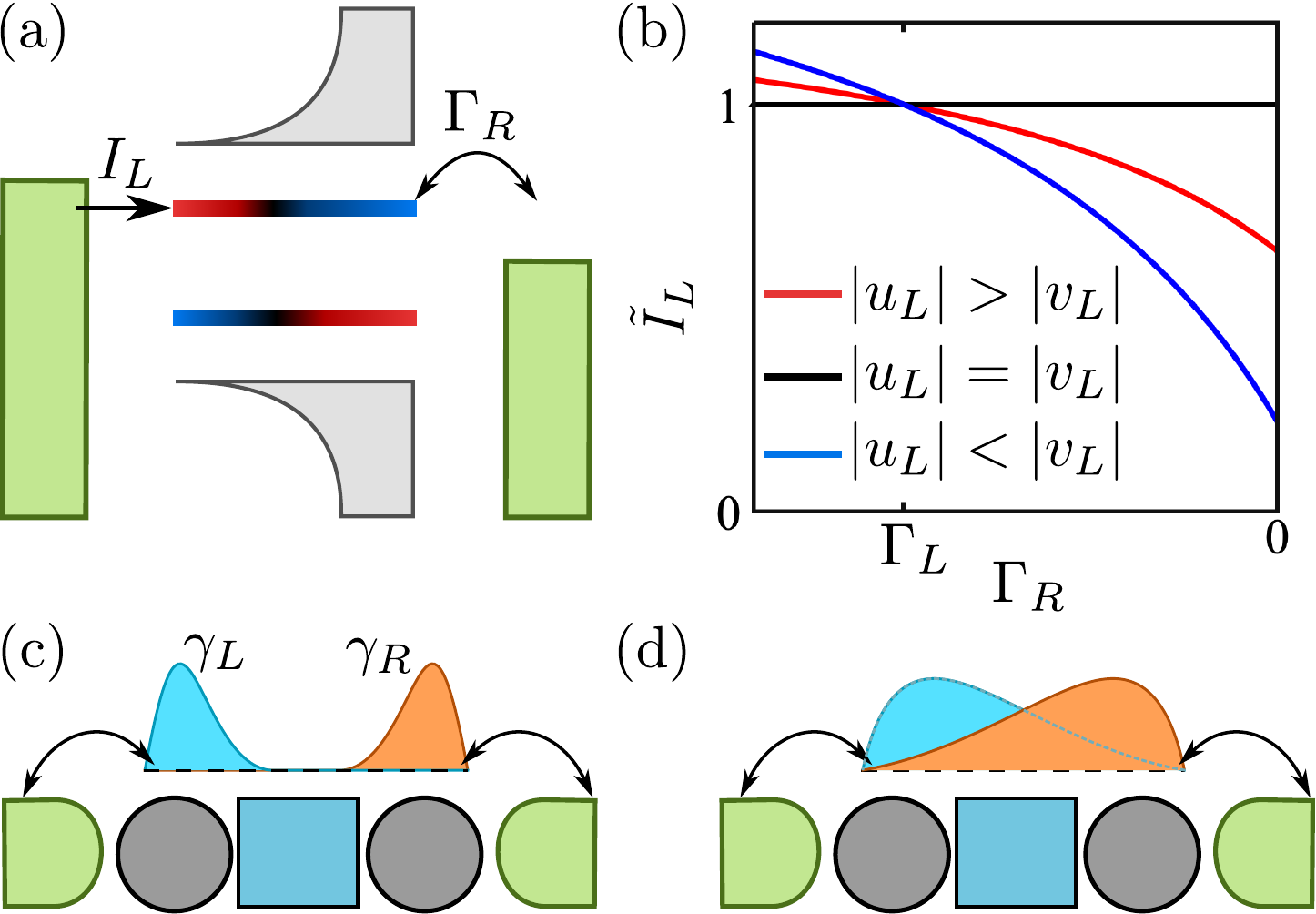}
\caption{(a) Sketch of the system considered, where a grounded superconductor couples to 2 metallic electrodes. The applied bias voltage is such that transport occurs through one state, whose colors illustrate the local electron (red) and hole (blue) character. (b) Normalized current through the left lead, $\tilde{I}_L=I_L(\Gamma_R)/I_L(\Gamma_R=\Gamma_L)$ as the coupling to the right one is changed. Here, we use Eq.~\eqref{Eq:I_L_complete}, considering left-right symmetry ($u_L=u_R\equiv u$ and $u_R=v_R\equiv v$), and $u=2v$, $u=v$, and $v=4u$ for red, black, and blue curves. (c) and (d) Minimal Kitaev chains at the sweet spot (c) and away from it (d). These two situations correspond to local $u=v$ and $u\neq v$ at the ends, respectively.}
\label{Fig1}
\end{figure}

In this work, we propose a method for measuring the local coherence factors of a subgap state in a grounded superconductor using the current through normal electrodes, Fig.~\ref{Fig1}(a). For locally symmetric coherence factors $|u|=|v|$, the current becomes independent of the coupling to any other lead. Therefore, the current behavior as a function of the tunnel rate to another lead can be used to find spots with local electron-hole symmetry, $|u|=|v|$, Fig.~\ref{Fig1}(b). Sequential tunneling spectroscopy measurements have been proposed in Coulomb-blockaded Majorana islands~\cite{Hansen2018probing}, where the conductance across even-odd charge transitions provides information about the coherence factors at both ends of the island. In contrast, our approach works for grounded superconductors and uses the current behavior as a function of the lead couplings to infer the local coherence factors. This provides spatially resolved information and complements non-local conductance measurements~\cite{danon2020nonlocal,Menard_PRL2020,Pikulin_arXiv2021,Poschl_PRB2022,Aghaee_PRB2023}, which probe the product of the coherence factors at both ends.

We use proposed coherence factor measurements to quantify the spatial separation between MBSs in short artificial Kitaev chains.  To this end, a measurement of the coherence factors can be used to estimate the Majorana polarization (MP)~\cite{sticlet2012spin, sedlmayr2015visualizing, sedlmayr2016majorana, aksenov2020strong, tsintzis2022creating,Awoga_PRB2024,Souto_PRB2025}. Non-overlapping MBSs feature $|u|=|v|$, while their overlap modifies these coherence factors, Figs.~\ref{Fig1}(c, d). Therefore, the current behavior as a function of the coupling to a second electrode at the other end of the chain allows us to distinguish between non-overlapping and overlapping MBSs and to estimate the local MP. We derive an analytic expression for the MP as a function of 2 current measurements. We benchmark this expression against numerical calculations for 2- and 3-site Kitaev chains, that include many-body interactions and finite Zeeman splittings, showing that our method provides a good approximate measurement for MP for a wide range of parameters. In combination, these results suggest that our proposal could be used to determine the degree of localization of MBSs and optimize sweet spots using, {\it e.g.} machine learning methods~\cite{Koch_PRAp2023, van2024cross, Benestad_PRB2024}.

The paper is organized as follows. In Sec.~\ref{idea_section}, we derive an expression relating the current and MP which is valid for a single subgap state in the limit of weak tunnel coupling. In Sec.~(\ref{theory section}), we describe the Kitaev chain and one of its practical realizations, a microscopic model describing QDs coupled via superconducting segments. In addition, we introduce the quality factors used to characterize sweet spots. In Sec.~(\ref{kitaev section}) we provide analytical and numerical results for the 2-site Kitaev chain. We expand these results for the microscopic model for 2 and 3-site Kitaev chains in Secs.~(\ref{Microscopic2site}) and (\ref{3siteSection}), respectively. We present our concluding remarks in Sec.~(\ref{conclusions}). Throughout the paper, we use units where $e=h=k_B=1$.

\section{Measuring coherence factors} \label{idea_section}

We start our analysis by studying the currents through a subgap state in a superconductor that couples to two metallic electrodes, as sketched in Fig.~\ref{Fig1}(a). We assume that the state is spinless, well-separated from other states and ignore interactions, although some of these assumptions will be relaxed in the next section. States in superconductors are characterized by the coherence factors, $u$ and $v$, that are generally position-dependent. These coefficients determine the local particle/hole-like character of states and their response to local electric fields. For transport, it is important to introduce the local coherence factors at the positions where the state couples to the leads, $u_{\nu}$ and $v_\nu$, with $\nu=L,R$ denoting the left/right side.

The Hamiltonian describing the subgap state is given by
\begin{equation}
    H_s=\xi \alpha^\dagger \alpha\,,
\end{equation}
where $\xi$ is the energy of the state and $\alpha$ ($\alpha^\dagger$) is the operator annihilating (creating) a quasiparticle excitation in the state, $\ket{1} = \alpha^\dagger \ket{0}$ ($\alpha \ket{0} = 0$). The tunnelling to the leads is given by
\begin{equation}
    H_T=\sum_{\nu k}t_\nu f^\dagger_{\nu k}\left(u_{\nu} \alpha + v_\nu \alpha^\dagger  \right)+\mbox{H.c.}\,,
\end{equation}
where $t$ is the tunneling amplitude and $f^\dagger_{\nu\sigma}$ creates an electron with momentum $k$ in lead $\nu$.

The local current has a very compact expression in the limit $V \gg T \gg \Gamma_{L, R}$, where $\Gamma_\nu=2\pi\rho_\nu t^{2}_\nu$ is the tunnel rate, $\rho$ the density of states in the leads at the Fermi level, $T$ the temperature, and $V$ the applied bias voltage. For $V_L = -V_R = V/2$ and $|V|/2\gg|\xi|$, we obtain (see App.~\ref{AppCoherenceF} for the derivation)

\begin{equation}
\label{Eq:I_L_complete}
    I_L = \Gamma_L \frac{2 \Gamma_L |u_L|^2 |v_L|^2 + \Gamma_R \left( |u_L|^2 |u_R|^2 + |v_L|^2 |v_R|^2 \right)}{\Gamma_L \left(|u_L|^2 + |v_L|^2 \right) + \Gamma_R \left(|u_R|^2 + |v_R|^2 \right)}.
\end{equation}
Interestingly, for $|u_L|=|v_L|$, the local current simplifies to $I_L=\Gamma_L|u_L|^2$, becoming independent of the right coherence factors and tunnel rates. Physically, this is due to the current being independent from the occupation of the state and, therefore, to the right current. In general, the dependence of the local current on the tunnel rates can be used to extract the local coherence factors. This is better seen when taking the $\Gamma_R = 0$ limit, 
\begin{equation} \label{ILGRZero}
    I_L (\Gamma_R = 0) = \Gamma_L \frac{2 |u_L|^2 |v_L|^2}{|u_L|^2 + |v_L|^2}.
\end{equation}
The symmetrically coupled case, $\Gamma_R=\Gamma_L$, is also interesting, where
\begin{equation}
    I_L (\Gamma_R = \Gamma_L) = \Gamma_L \frac{2 |u_L|^2 |v_L|^2 + |u_L|^2 |u_R|^2 + |v_L|^2 |v_R|^2}{|u_L|^2 + |v_L|^2 + |u_R|^2 + |v_R|^2}.
\end{equation}
Considering  $|u_R| = |u_L| = |u|$ and $|v_R| = |v_L| = |v|$, the ratio between the currents is

\begin{equation} \label{ILSpinlessMain}
    \frac{I_L (\Gamma_R = 0)}{I_L (\Gamma_R = \Gamma_L)} = \left(\frac{2 |u| |v|}{|u|^2 + |v|^2} \right)^2.
\end{equation}
As we will see in the following section, this ratio is related to the MP. We then benchmark the developed relation between current and MP against numerical calculations of quantum transport through artificial Kitaev chains implemented in arrays of QDs. Our model includes additional states, the spin degree of freedom, and many-body interactions, testing the validity of our expressions beyond the hypothesis in this section.

\section{Artificial Kitaev chains} \label{theory section}
Short Kitaev chains are examples of systems where the properties of subgap states can be locally controlled, leading to many different scenarios, including: Majorana states (spinless states with $|u|=|v|$ and $\xi=0$), particle-hole symmetric states ($|u|=|v|$), and more generic states ($|u|\neq |v|$).

\subsection{Model}
We start by describing the microscopic model that leads to effective Kitaev chains~\cite{tsintzis2022creating}. The system consists of normal QDs coupled via narrow superconductors that host subgap states, modeled as QDs with induced superconducting correlations. We consider an external magnetic field that polarizes the spins in the QDs. The Hamiltonian that describes the QDs is given by

\begin{equation}
\begin{split}
    H_{\rm QDs} =& \sum_{i, \sigma}^N (\varepsilon_{i} + s_\sigma V_{z, i}) n_{i, \sigma} +\sum_i^N U_i n_{i, \uparrow}n_{i, \downarrow}+\\
    &\sum_{i}^{n_s}\left(\Delta c_{2i, \uparrow}^\dagger c_{2i, \downarrow}^\dagger + {\rm H.c.} \right),
    \label{Eq:Hmicroscopic}
\end{split}
\end{equation}
which describes an effective Kitaev chain with $n_s=(N+1)/2$ sites, $N$ being an odd number. Here, $\varepsilon_{i}$ is the on-site energy of QD $i$, $V_z$ is the Zeeman energy ($s_{\uparrow/\downarrow} = \pm 1$), and  $U_i$ is the local electrostatic repulsion, with $n_{i,\sigma}=c^\dagger_{i,\sigma} c_{i,\sigma}$ being the number operator and $c_{i,\sigma}$ the annihilation operator for electrons with spin $\sigma=\uparrow,\downarrow$ at site $i$. We describe the coupling between QDs and the superconductor as an induced pairing amplitude $\Delta$.

The tunneling between the QDs is given by
\begin{equation} \label{TunnelingQDs}
    H_T= \sum_{i, \sigma}^{N-1}\left[ t_i c_{i+1, \sigma}^\dagger c_{i, \sigma} + t_{i}^{so} s_{\sigma}c_{i+1, \sigma}^\dagger c_{i, \bar{\sigma}}+ {\rm H.c.} \right],
\end{equation}
where $t$ and $t^{\mathrm{so}}$ are the spin-conserving and spin-flip tunneling amplitudes, respectively, with $t^{\mathrm{so}}$ arising from spin-orbit coupling and $\bar{\sigma}$ denoting the spin opposite to $\sigma$. The leads are described by
\begin{equation}
    H_l=\sum_{\nu k\sigma}\xi_{\nu, k \sigma}f_{\nu k \sigma}^\dagger f_{\nu, k \sigma}\,,
\end{equation}
where $\nu=L$ ($R$) labels the left (right) lead. The coupling between the system and the leads is given by
\begin{equation}
    H_T^{l}=\sum_{k\sigma}\left(t_L\,f_{L, k \sigma}^\dagger c_{1,\sigma}+t_R\,f_{R, k \sigma}^\dagger c_{N,\sigma}+{\rm H.c.}\right)\,,
\end{equation}
where $N$ is the last site of the chain and we have considered the spectrum of the lead and the tunnel couplings to be spin-independent for simplicity. The total Hamiltonian is 
\begin{equation} \label{microscopicH}
    H = H_{\rm QDs} + H_T+H_l+H_{T}^l.
\end{equation}
We calculate the current through the system using a rate equation approach, valid in the limit of weak tunnel couplings between the leads and the system~\cite{kirvsanskas2017qmeq}.

In the weak coupling limit between the QDs, $t_i, t_i^{so} \ll \Delta$, it is possible to integrate the superconducting QDs out of the problem. The ratio between ECT ($\tau_i$) and CAR ($\delta_i$) amplitudes between the normal QDs is controlled by $\varepsilon_i$, with $i$ even~\cite{liu2022tunable, tsintzis2022creating}, and these two amplitudes represent the hoppings and p-wave pairing amplitudes in an effective Kitaev chain~\cite{liu2022tunable}. For sufficiently large spin splitting, the effective Hamiltonian emulates a Kitaev chain~\cite{Kitaev2001}
\begin{equation} \label{Kitaev2 Hamiltonian}
    H_K = -\sum_{i=1}^{n_s} \mu_i d_i^\dagger d_i +  \sum_{i=1}^{n_s-1}\left(\tau_i d_{i+1}^\dagger d_i + \delta_i d_i^\dagger d_{i+1}^\dagger + {\rm H.c.} \right),
\end{equation}
where $d_i$ are combinations of $c_{i,\uparrow}$ and $c_{i, \downarrow}$ and $\mu_i$ is the on-site chemical potential. In this approximation, the current can be calculated exactly using the scattering matrix approach. In the following, we will refer to the model described in Eq.~\eqref{microscopicH} as the microscopic model and the one given in Eq.~\eqref{Kitaev2 Hamiltonian} as the Kitaev chain. We will focus the majority of our analysis on the 2-site Kitaev chain, Eq.~(\ref{Kitaev2 Hamiltonian}), and its microscopic model, Eqs.~(\ref{Eq:Hmicroscopic}-\ref{microscopicH}) with $N = 3$, and discuss results for the microscopic realization of 3-site chains ($N = 5$) in Sec.~\ref{3siteSection}.

In short Kitaev chains, there is no topological phase. Nevertheless, localized MBSs can appear at discrete sweet spots in parameter space.
The local overlap between MBSs can be quantified by the MP~\cite{sticlet2012spin, sedlmayr2015visualizing,tsintzis2022creating, dourado2025twositekitaevsweetspots}

\begin{equation} \label{MPsigma}
    M_i= \frac{\sum_\sigma w_{i, \sigma}^2 - z_{i, \sigma}^2}{\sum_\sigma w_{i, \sigma}^2 + z_{i, \sigma}^2} ,
\end{equation}
where $ w_{i, \sigma} = \bra{O}\left( c_{i,\sigma} + c_{i, \sigma}^\dagger \right)\ket{E}$ and $z_{i, \sigma} = \bra{O} \left( c_{i, \sigma} - c_{i, \sigma}^\dagger \right) \ket{E}$, with $\ket{E,O}$ being the even/odd parity ground state. In the non-interacting limit~\cite{Awoga_PRB2024}, 
\begin{equation} \label{MPsigma_2}
    M_i= \frac{\sum_{ \sigma}2 u_{i, \sigma}v_{i, \sigma}}{\sum_{\sigma} u_{i, \sigma}^2 + v_{i, \sigma}^2} ,
\end{equation}
where $u_{i, \sigma}$ and $v_{i, \sigma}$ are the local coherence factors.

The MP can be estimated using the local current when varying the coupling to the other lead. We define the estimated MP as 
\begin{equation} \label{measuredMP}
    \mathcal{M}_\nu = \sqrt{\frac{I_\nu (\Gamma_{\Bar{\nu}} = 0)}{I_\nu (\Gamma_{\Bar{\nu}} = \Gamma_\nu)}}, 
\end{equation}
where $\nu\in \{L, R\}$ and $\bar{\nu}\neq\nu$  represents the opposite end of the system. Comparing Eqs.~(\ref{ILSpinlessMain}) and (\ref{MPsigma}) directly shows $|M| = \mathcal{M}$ for the spinless system with left-right symmetry ($u_L=u_R=u$ and $v_L=v_R=v$). This shows that 2 current measurements are enough to estimate the local coherence factors in the symmetric case, which can be used for experimentally measuring the MP. Although in what follows we focus on Eq.~(\ref{measuredMP}) to estimate the MP, other regimes with different $\Gamma_\nu$ values can be used to this end, see App.~\ref{AppCoherenceF}. 

To reach a sweet spot in 2-site chains, we impose three conditions: (i) degenerate ground states; (ii) maximal MP at the system ends; and (iii) a finite gap to the excited states. In the effective 2-site Kitaev chain, this requirement translates into $\tau = \delta$. Additionally, the site energies in the Kitaev chain should be aligned with the chemical potential of the superconductor, i.e. $\mu_1 = \mu_2 = 0$~\cite{Leijnse2012}. These conditions enable the emergence of fully localized MBSs on each site, $|M_1| = |M_2| = 1$, and degeneracy between the  even and odd ground states, $\delta E_0 = 0$. In the microscopic model, Eqs.~(\ref{Eq:Hmicroscopic}- \ref{TunnelingQDs}), it is possible to find sweet spots for $V_Z\gtrsim\Delta$, with most of the MBS wavefunctions localized at the outer QDs ($|M| \lesssim 1$), while ensuring degeneracy of the ground state~\cite{liu2022tunable,tsintzis2022creating}.

\subsection{2-site Kitaev chain} \label{kitaev section}

We first analyze the two-site Kitaev chain with symmetric on-site energies, $\mu\equiv\mu_1=\mu_2$, Eq.~(\ref{Kitaev2 Hamiltonian}), although the presented conclusions hold for asymmetric setups, see App.~\ref{AppGLR}. 

\begin{figure}
\centering
\includegraphics[width=\linewidth]{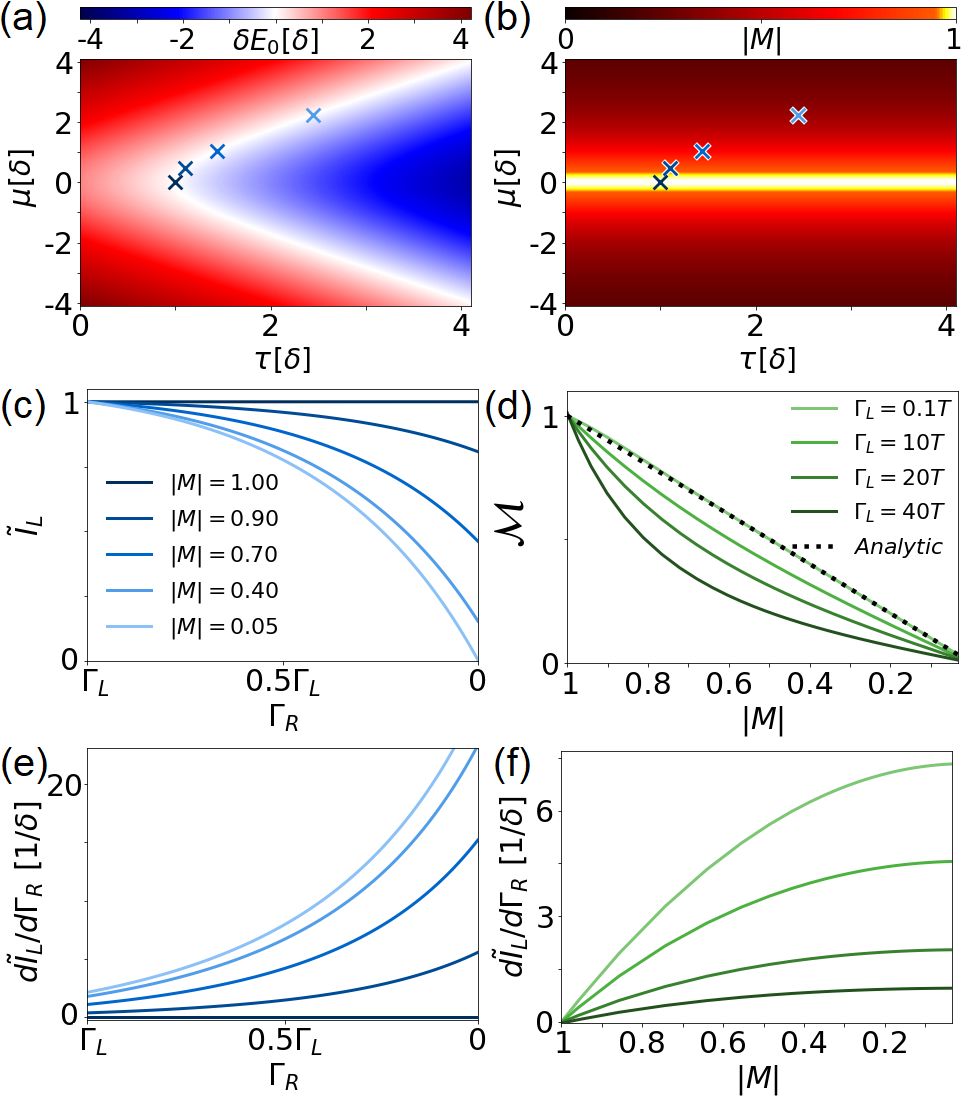}
\caption{Results for the 2-site Kitaev chain. (a)-(b) $\delta E_0$ and $|M|$ as functions of $\tau$ and $\mu$, both parameters expressed in units of $\delta$. (c) Normalized current, $\Tilde{I}_{L} = I_{L}/I_{L}(\Gamma_R = \Gamma_L)$, as a function of $\Gamma_R$ for different MPs. (d) Estimated MP based on Eq.~(\ref{measuredMP}), as a function of the MP calculated from the definition in Eq.~(\ref{MP 2site Kitaev}), for several values of $\Gamma_L$. (e)-(f) $d\Tilde{I}_{L}/d\Gamma_R$ as a function of $\Gamma_R$ and $|M|$, respectively. Parameters: $T = \delta/62$, $V_L = - V_R = 0.05\delta$, in (c) $\Gamma_L = 0.1 \delta$, and in (f) $\Gamma_R = 0.5\Gamma_L$.}
\label{Kitaev2}
\end{figure}

The energy splitting between the even and odd fermion-parity ground states is $\delta E_0 = E_0^{odd} - E_0^{even}=\sqrt{1 + \mu^2}-|\tau|$ ($\mu$ and $\tau$ are represented in units of $\delta$), where positive (negative) values indicate that the lowest energy state has an even (odd) fermion parity. This is shown in Fig.~\ref{Kitaev2}(a), where $\delta E_0$ is calculated as a function of $\tau$ and $\mu$. The ground state degeneracy, $\delta E_0=0$, is represented by the white lines in Fig.~\ref{Kitaev2}(a), which mark transitions between even and odd ground states. We calculate the MP by applying Eq.~(\ref{MPsigma}), where the coherence factors are (up to a normalization factor) $u_{1 (2)} = -\mu -\sqrt{1 +\mu^2}$ and $v_{1(2)} = \mp 1$, which for this model simplifies to ($|M_1| = |M_2| = |M|$)~\cite{Tsintzis2024}
\begin{equation} \label{MP 2site Kitaev}
    |M|= \frac{1}{\sqrt{1 + \mu^2}}.
\end{equation}
Interestingly, for $\mu_1 = \mu_2$, the MP does not depend on the ratio $\tau/\delta$. The color scale in Fig.~\ref{Kitaev2}(b) shows that deviations from $\mu = 0$ reduce the MP for any $\tau$. The conditions $\delta E_0 = 0$ and $|M| = 1$ are fulfilled at discrete points in parameter space, $\mu_1=\mu_2=0$ and $|\tau| = 1$. In the following, we compare the current at and away from the sweet spot. As the ground state splitting can be measured via local spectroscopy, we initially focus on the situation where the ground state is degenerate, $\mu^2=\tau^2-1$.

For $|M| < 1$, the two MBSs connect to both leads, causing the currents to depend on both tunneling rates, $\Gamma_{L,R}$. To show this, we select points in parameter space where $\delta E_0 = 0$, but corresponding to different values for the MP. These points are marked by the blue crosses in Figs.~\ref{Kitaev2}(a, b) and match the color of the curves in Figs.~\ref{Kitaev2}(c, e). Here, $I_L$ is calculated using the scattering matrix approach with no particular parameter restriction, and its dependence on $\Gamma_R$ is explored, see App.~\ref{AnalyticalGLL} for details. At the sweet spot, where $|M|=1$, $I_{L}$ does not depend on $\Gamma_R$, dark blue line in Fig.~\ref{Kitaev2}(c). Lower values of $|M|$ result in a current reduction when decreasing the coupling to the right lead,  $\Gamma_R\to 0$. The amplitude of the current suppression allows us to assess the MP.

In the limit $\Gamma_{L,R}\ll \,T\ll\delta$, we calculate the current ratio $I_L(\Gamma_R=0)/I_L(\Gamma_R = \Gamma_L)$ for the Kitaev chain model. 
Equation~(\ref{measuredMP}) provides an estimation for MP,  as shown in the lightest green and dotted lines in Fig.~\ref{Kitaev2}(d), where we considered a small voltage bias in comparison to the excitation gap. Increasing the coupling to the electrodes leads to deviations from the analytic result. Nevertheless, $I_L(\Gamma_R = 0)/I_L(\Gamma_R = \Gamma_L)$ peaks for $|M|=1$ for any value of $\Gamma_L$. Therefore, the local measurements can still be used to maximize MP, where in experiments, the flatness of $I_L$ with respect to $\Gamma_R$ represents well separated MBSs. Increasing $V$ makes the measurement of MP independent of $\Gamma_{L}$ and $T$, as discussed in App.~\ref{GeneralNuLLKitaev}.

Our method to estimate and maximize MP can be summarized in measuring $I_{L}$ for $\Gamma_R = \Gamma$ and $\Gamma_R = \Gamma - \delta \Gamma_R$. Increasing $\delta \Gamma_R$, improves the difference between the signals for high and low MP points. We emphasize that no fine-tuning is required in the choice of the parameters, including $\Gamma$ and $\delta \Gamma_R$. Figure~\ref{Kitaev2}(e) illustrates this point, showing the derivative of the current $d\Tilde{I}_{L}/d\Gamma_R$ for different values of $\Gamma_R$, where values different from zero indicate overlapping MBSs at the ends of the system. The dependence on the MP is better illustrated in Fig.~\ref{Kitaev2}(f), where $d \Tilde{I}_L/d\Gamma_R$ increases as $|M|$ diminishes.

\subsection{Microscopic Model} \label{Microscopic2site}

\begin{figure}
\centering
\includegraphics[width=\linewidth]{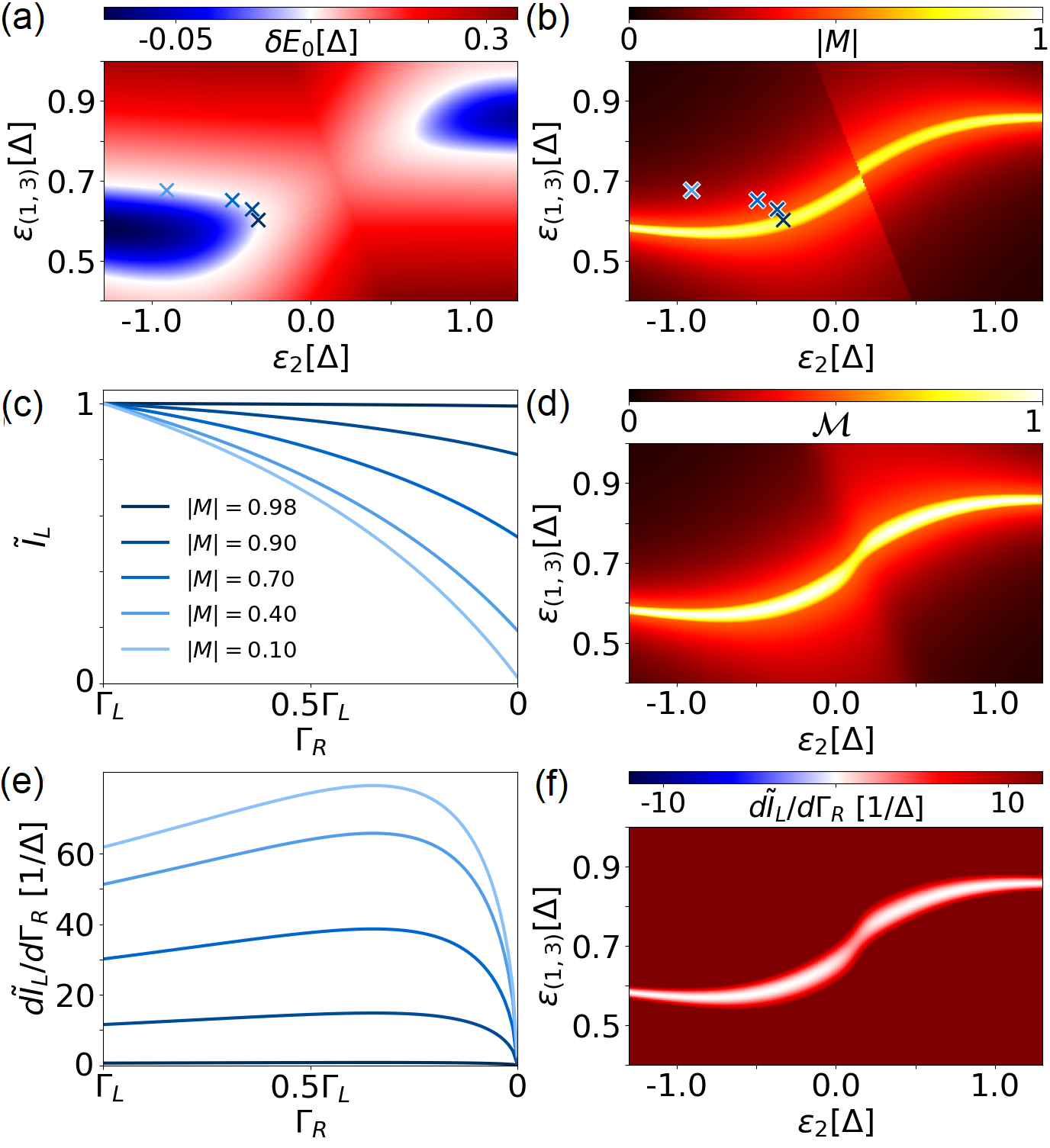}
\caption{Results for the microscopic model of an effective 2-site Kitaev chain. (a)-(b) $\delta E_0$ and $|M|$ as functions of $\varepsilon_{1, 3} = \varepsilon_1 = \varepsilon_3$ and $\varepsilon_2$. (c) $\Tilde{I}_L$ as a function $\Gamma_R$ for several values of $|M|$. (d) $\mathcal{M}$, see Eq.~(\ref{measuredMP}), as a function of $\varepsilon_{1, 3}$ and $\varepsilon_2$, closely resembling the MP shown in (b). (e) $d\Tilde{I}_{L}/d \Gamma_R$ as a function of $\Gamma_R$ for the same values of $|M|$ as in (c). (f) $d\Tilde{I}_{L}/d \Gamma_R|_{\Gamma_R = \Gamma_L}$  as a function of the QD levels, also reproducing the MP similar to (b) and (d). Parameters: $V_z = 2.5\Delta$, $t_i = 0.5\Delta$, $t^{so}_i = 0.2t_i$, $U_i = 5\Delta$, $T = \Delta/62$, $\Gamma_L = T/10$, and $V_L = -V_R = 0.01\Delta$.}
\label{SpinfulKitaev2}
\end{figure}

We now study the microscopic model, introduced in Eqs.~(\ref{Eq:Hmicroscopic}-\ref{microscopicH}), that considers both spin channels and electron-electron interactions in the QDs~\cite{tsintzis2022creating}. For sufficiently large Zeeman splitting, $V_Z > \Delta$, the system typically features two sweet spots, characterized by a zero energy splitting between the even and odd parity ground states ($\delta E_0=0$), and high MP. Similar to the previous section, we show the energy splittings and the MP as functions of the parameters, which for the microscopic model are the QD levels, in Figs.~\ref{SpinfulKitaev2}(a, b), respectively. For the parameters in Fig.~\ref{SpinfulKitaev2}, $|M| \approx 0.985$ at the sweet spot ($\delta E_0=0$), denoted by the dark blue cross.

Figure~\ref{SpinfulKitaev2}(c) shows the current $I_L$ as a function of $\Gamma_R$, similar to what we showed in Fig.~\ref{Kitaev2}(c) for the Kitaev chain.
We select points in a degeneracy line ($\delta E_0=0$), blue crosses in Figs.~\ref{SpinfulKitaev2}(a, b), corresponding to different values of $|M|$. As expected, $\tilde{I}_L$ is almost flat as a function of $\Gamma_R$ for $|M|\approx1$. Decreasing $|M|$ leads to an increased difference between the current $\tilde{I}_L$ measured for different $\Gamma_R$ values. We note that the quantitative agreement is not limited to the points with energy degeneracy in the limit $\delta E_0 < T$. This allows us to track MP by applying Eq.~(\ref{measuredMP}) for the full range of parameters, as shown in Fig.~\ref{SpinfulKitaev2}(d), where $\mathcal{M}$ accurately reproduces $|M|$ in Fig.~\ref{SpinfulKitaev2}(b), especially close to the high-MP regions. 

In Figs.~\ref{SpinfulKitaev2}(e, f), we consider small variations in $\Gamma_R$ and show the current derivative as a function of $\Gamma_R$, that corresponds to the limit $\delta\Gamma_R\to 0$. Similar to the Kitaev chain results, see Figs.~\ref{Kitaev2}(e, f), minimizing the value of $d \tilde{I}_L/d\Gamma_R$ maximizes the MP.

Therefore, the dependence of the local current on the coupling to the leads can be used for finding  Majorana sweet spots. In particular, well-localized MBSs result in a current that is independent to variations on the coupling with other leads, while overlapping MBSs coupling to the same lead are sensitive to these variations. Our proposal can be used in combination to other known methods such as the nonlocal conductance \cite{danon2020nonlocal, tsintzis2022creating}, the coupling of probe QDs to assess the MBS localization \cite{Deng_Science2016,Prada_PRB2017,Clarke_PRB2017,Seoane2023}, or machine learning tuning protocols \cite{van2024cross, Benestad_PRB2024}, as our method does not require moving the system from the sweet spot to evaluate the MP and can be used as a post-tuning verification of the sweet spot conditions.

\subsection{Dependence on the pin-splitting} \label{weakVz}

\begin{figure}
\centering
\includegraphics[width=\linewidth]{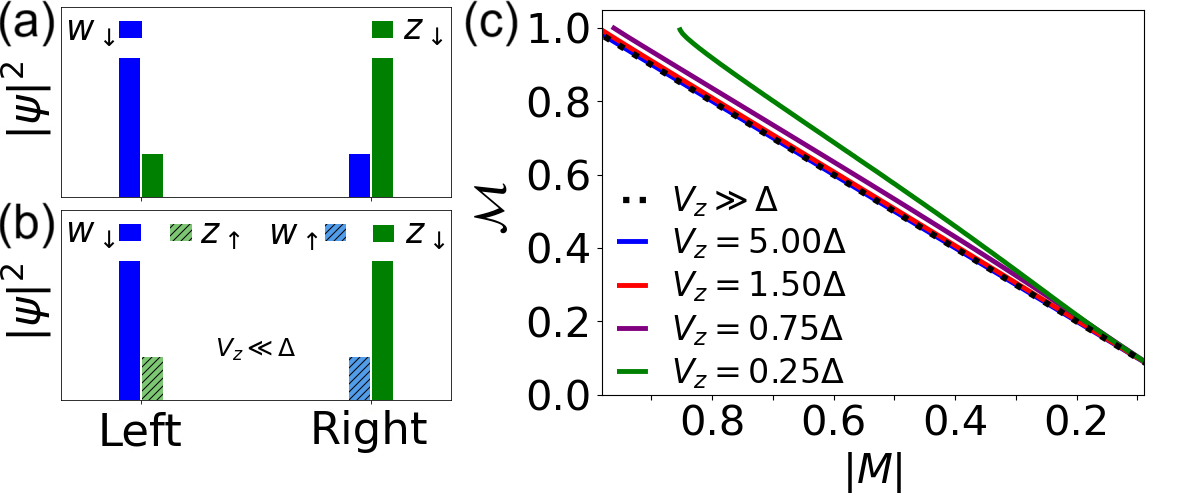}
\caption{(a) Overlapping Majorana wavefunctions at the left and right QDs for the microscopic model and strong polarization ($V_z \gg \Delta$). (b) Similar to (a) but for weak magnetic fields, where the reduction to the MP comes from the spin-up components. (c) $\mathcal{M}$ as a function of $|M|$ for different values of $V_z$. Parameters are the same as in Fig.~\ref{SpinfulKitaev2}.}
\label{weakVzMain}
\end{figure}

We now investigate the accuracy of Eq.~(\ref{measuredMP}) for estimating MP at low magnetic fields. We find that our method provides a precise measurement of the MP in the regime $V_z \geq \Delta$. In this case, one of the spin channels is gaped out and deviations from the ideal MP value ($|M|=1$) can mostly come from overlapping Majorana states with the same spin, as depicted in Fig.~\ref{weakVzMain}(a).

Our method can detect MP reductions that come from MBSs with the same spin overlapping, schematically represented in Fig.~\ref{weakVzMain}(a). Another mechanism to reduce MP consists at lowering the applied magnetic field, allowing the ground state wavefunctions to have some weight on the MBS components with the opposite spin. This case is schematically represented in Fig.~\ref{weakVzMain}(b), where both $w_\downarrow^2$ and $z_\uparrow^2$ are finite, leading to $|M|<1$ according to Eq.~\eqref{MPsigma}. In contrast, the two spin channels contribute independently to the current, and their ratio yields, see App.~\ref{AppCoherenceF}, 
\begin{equation} \label{ILSpinfulMain}
    \frac{I_L (\Gamma_R = 0)}{I_L(\Gamma_R = \Gamma_L)} = \frac{4}{\sum_\sigma |u_\sigma|^2 + |v_\sigma|^2} \sum_\sigma \frac{|u_\sigma|^2 |v_\sigma|^2}{|u_\sigma|^2 + |v_\sigma|^2},
\end{equation}
which cannot be related straightforwardly to the MP as defined in Eq.~(\ref{MPsigma_2}). 

From Eq.~\eqref{ILSpinfulMain}, it becomes clear that the current ratio matches the MP expression for a vanishing weight in one of the spin species, {\it i.e.} $V_z\to\infty$. To get insight in the range of validity of the spinless approximation, we have simulated the effective 2-site Kitaev chain using the microscopic model, Eq.~\eqref{microscopicH} (N = 3), for different $V_z$ values, Fig.~\ref{weakVzMain}(c). The estimated MP, $\mathcal{M}$, almost overlaps with $|M|$ for $V_z\gg\Delta$ (dotted line). The deviation from the real MP value is $\sim 3 \%$ for $V_z = 0.75 \Delta$ and $\sim 14 \%$ for $V_z = 0.25 \Delta$. We note that Coulomb interactions reduces the difference between the measured and real MP. As pointed out in Ref.~\cite{tsintzis2022creating}, the maximum value for the MP increases with $U$. Therefore, the errors described above can be significantly reduced for $U > 5\Delta$. In current experiments, the Zeeman splitting can be larger than the gap~\cite{Dvir2023, bordin2024signatures, ten2024two}, making our proposal viable and precise as a measurement for the MP. In any case, we remark that the ratio between the currents is a good quantity to optimize MP:  $\mathcal{M}$ peaks at the maximum value of $|M|$ for any $V_z$, Fig.~\ref{weakVzMain}(c).

\subsection{3-site Kitaev chain} \label{3siteSection}

\begin{figure}
\centering
\includegraphics[width=\linewidth]{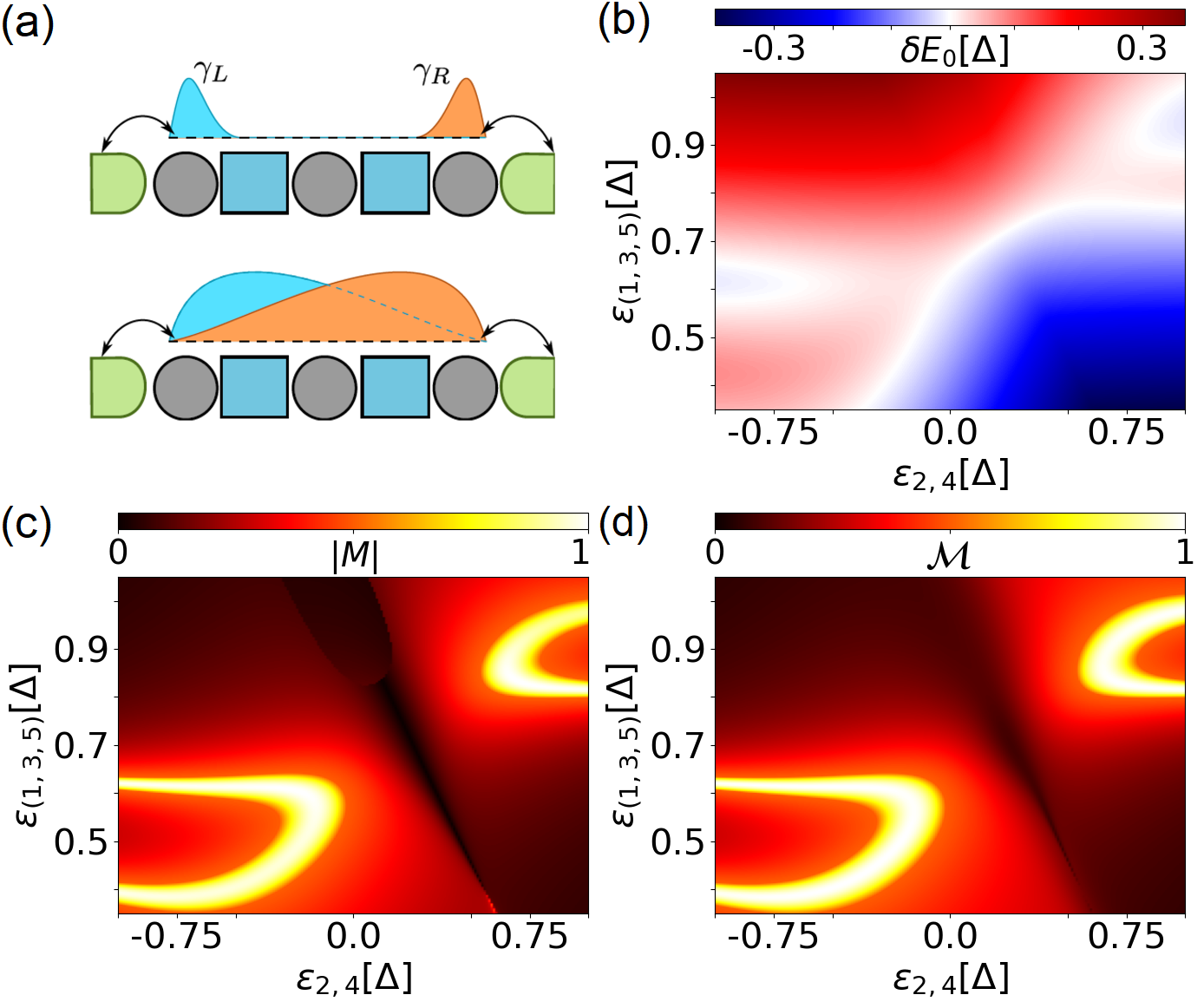}
\caption{(a) Results for the microscopic model for an effective 3-site Kitaev chain in a 5-QD setup, where the grey and blue regions represent normal and superconducting QDs, respectively. The MBS wavefunctions are pictorially represented in blue and orange. The levels of the QDs are controlled via external gates $\varepsilon_i$. (b)-(c) $\delta E$ and $|M|$ as functions of the QD levels $\varepsilon_{1, 3, 5}$ varied together, and also the level of the superconducting QDs $\varepsilon_{2, 4}$. (d) $\mathcal{M}$ as a function of the QD levels reproducing the MP in (c).}
\label{3siteFullmodel}
\end{figure}

We now test our method in longer chains. Increasing the number of sites in Kitaev chains is an important step towards improving Majorana robustness and eventually achieving topological protection~\cite{bordin2024signatures, haaf2024edgebulkstatesthreesite, dourado2025twositekitaevsweetspots}. Recently, it has been demonstrated that going from 2 to 3-site Kitaev chains already increases the robustness of zero-bias peaks against variations of parameters~\cite{bordin2024signatures,haaf2024edgebulkstatesthreesite}. Here, we consider the extended array of QDs in the $V_z\gg\Delta$ limit, shown in Fig.~\ref{3siteFullmodel}(a), which emulates a 3-site Kitaev chain.

First, we identify the sweet spots as a function of the onsite energies of the normal, $\varepsilon_{1, 3, 5} = \varepsilon_1 = \varepsilon_3 = \varepsilon_5$, and superconducting QDs, $\varepsilon_{2, 4} = \varepsilon_2 = \varepsilon_4$. For that, we calculate both $\delta E_0$ and $|M|$, Figs.~\ref{3siteFullmodel}(b, c). We note that compared to the 2-site chain, there is an additional degeneracy line, which intercepts the left-lower $|M| \approx 1$ arc, giving rise to additional sweet spots~\cite{Dourado_Kitaev3}.

Our two-current measurement provides a direct mapping of MP for longer Kitaev chains, as $\mathcal{M}$ reproduces the MP map, see comparison between Fig.~\ref{3siteFullmodel}(d) and Fig.~\ref{3siteFullmodel}(c). Tuning into 3-site Kitaev chain sweet spots comprises additional challenges in contrast to the 2-site chain. This is mainly due to QD $3$ being coupled to $2$ superconductors, see Fig.~\ref{3siteFullmodel}(a), thus experiencing a different renormalization than the outermost QDs. This configuration can lead to the misalignment of the renormalized QD levels, causing a variety of sweet spots, with different degrees of robustness and localization, to emerge~\cite{Dourado_Kitaev3}. Regardless of this increased complexity in the 3-site chain, $\mathcal{M}$ is close to $|M|$ for each type of sweet spot.

\section{Conclusions} \label{conclusions}

In this work, we have proposed a way to infer the local coherence factors of subgap states in superconductors, based on local current measurements. Our method is based on a simple observation: the local current through a state with symmetric coherence factors, $|u|=|v|$, is independent from its occupation and, therefore, independent from its coupling to a second lead. We derive analytic expressions for the dependence of current on the coupling to a second lead and show that it can be used to infer the relation between $u$ and $v$.

For a short Kitaev chain, the measurement of the coherence factors provides an estimate for the local overlap between the MBSs, which is quantified by the MP. We derived analytical expressions for the relation between MP and quantity that depends only on normal current measurements. Our proposed measurements provide a good estimate for MP, also for a microscopic model that includes a finite magnetic field and electron-electron interactions. This provides a route for quantitatively determining the overlap between MBSs in short Kitaev chains.

A related work by M. Alvarado et al., Ref.~\cite{Alvarado_arXiv20205}, submitted in parallel, proposes extracting the Majorana polarization from the relative heights of conductance peaks that appear when probing a Kitaev chain with an Andreev bound state.

\section{Acknowledgments}
The authors thank Viktor Svensson for useful discussions. R.A.D. acknowledges financial support from the Conselho Nacional de Desenvolvimento Científico e Tecnológico (CNPq), Brazil (Grant No. 141461/2021-7) and Coordenação de Aperfeiçoamento de Pessoal de Nível Superior (CAPES), Brazil (Grant No. 88887.111639/2025-00). M.L. acknowledges funding from the European Research Council (ERC) under the European Unions Horizon 2020 research and innovation programme under Grant Agreement No. 856526, the Swedish Research Council under Grant Agreement No. 2020-03412, and NanoLund. R.S.S acknowledges funding from the Horizon Europe Framework Program of the European Commission through the European Innovation Council Pathfinder Grant No. 101115315 (QuKiT), the Spanish Comunidad de Madrid (CM) ``Talento Program'' (Project No. 2022-T1/IND-24070), and the Spanish Ministry of Science, innovation, and Universities through Grants PID2022-140552NA-I00 and CEX2024-001445-S.
\newpage
\appendix

\section{Transport calculations for the Kitaev chain} \label{AnalyticalGLL}

For the Kitaev chain, we obtain the current and conductances through the scattering matrix (S-matrix) formalism. The S-matrix can be obtained using the general expression~\cite{nilsson2008splitting, maiani2022conductance}

\begin{equation} \label{smatrix}
    S = 1 - 2\pi i W^\dagger \frac{1}{E - \mathcal{H}_M + i \pi  W W^\dagger} W,
\end{equation}
where $\mathcal{H}_M$ describes the system and $W$ represents the coupling between the system and the leads.

Once we have calculated the S-matrix, we obtain the current $I_{L}$ using the below expression~\cite{maiani2022conductance}.

\begin{equation} \label{GL}
\begin{split}
    I_{L} &=  \int dE \left(2 A_L + T_{RL} + A_{RL} \right)  \Tilde{f}(V_L) +\\ &- \int dE \left(T_{LR} - A_{LR} \right) \Tilde{f}(V_R),
\end{split}
\end{equation}
where $\Tilde{f}(V_\alpha) = f(E - V_\alpha) - f(E)$, with $f(E)$ being the Fermi function, $A_L = |a_L|^2$ is associated with the Andreev reflection process, $A_{RL} = A_{LR} = |a_{RL}|^2$ and $T_{RL} = T_{LR} = |t_{RL}|^2$ with the CAR and electron tunneling processes, respectively. 

We start by casting the Kitaev Hamiltonian, Eq. (\ref{Kitaev2 Hamiltonian}), into the Bogoliubov-de Gennes representation. In the basis $\psi_{M} = \begin{pmatrix}
    d_1 & d_2 & d_1^\dagger & d_2^\dagger
\end{pmatrix}$, the Hamiltonian reads

\begin{equation} \label{HM}
    \mathcal{H}_M = \begin{pmatrix}
        -\mu_1 & \tau & 0 & \delta \\ \tau & -\mu_2 & -\delta & 0 \\ 0 & -\delta & \mu_1 & -\tau \\ \delta & 0 & -\tau & \mu_2
    \end{pmatrix}.
\end{equation}
We proceed similarly for the tunneling Hamiltonian,

\begin{equation}
    H_T = \sum_k \left[t_L d_1^\dagger f_{L, k} + t_R d_2^\dagger f_{R, k} + {\rm H.c.}\right],
\end{equation}
such that $H_T = \frac{1}{2} \psi_{M}^\dagger W \psi_{leads} + \frac{1}{2} \psi_{leads}^\dagger W^\dagger \psi_{M}$.

The coefficients that contribute to the current are $S_{21} = t_{RL}$, $S_{31} = a_{L}$, and $S_{41} = a_{RL}$. Numerically, we obtain the S-matrix and substitute the coefficients into Eq.~(\ref{GL}). We emphasize that within this framework, there is no parameter restriction, in contrast to the results derived analytically via the rate equations approach, see App.~\ref{AppCoherenceF}. We use the numerical results to benchmark the analytical expressions and explore the role of electron-electron interactions and finite magnetic fields.

\section{Finite energy and parameter dependence in the 2-site Kitaev chain} \label{GeneralNuLLKitaev}

In this Appendix, we explore how the results shown in Fig.~\ref{Kitaev2} change as a function of different parameters, including the energy splittings, temperature $(T)$, $\Gamma_{L, R}$, and $V_{L, R}$.

\begin{figure}
\centering
\includegraphics[width=\linewidth]{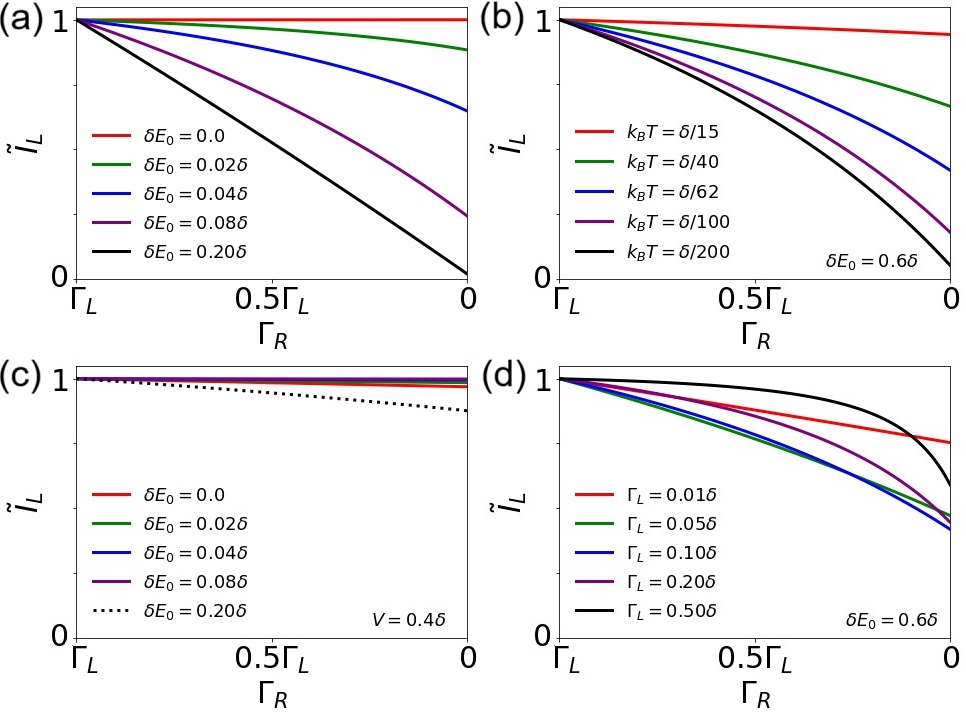}
\caption{Results for finite $\delta E_0$. (a) $\Tilde{I}_{L}$ a a function of $\Gamma_R$ for different values of $\delta E_0$. (b) For fixed $\delta E_0 = 0.06 \delta$ for different values of $T$. (c) Same as (a) but considering voltage bias $V_L = -V_R = 0.4\delta$. (d) $\Tilde{I}_{L}$ as a function of $\Gamma_R$ for several values of $\Gamma_L$ for fixed $\delta E_0 = 0.6 \delta$. Unless specified otherwise, we consider $T = \delta/62$, $V_L = -V_R = -0.01\delta$, and $\Gamma_L = 0.1\delta$.}
\label{energies}
\end{figure}

We start by setting $\mu = 0$, which ensures that $|M| = 1$, and we take several values for the hopping, as $\delta E_0$ becomes finite for $\tau > 1$ (in units of $\delta$). We plot the normalized current as a function of $\Gamma_R$ in Fig.~\ref{energies}(a). We show the sweet spot result ($\tau = 1$) with a red line, where no coupling between the left and right sides exist, such that $I_{L}$ features a plateau with respect to $\Gamma_R$. For $\delta E_0 \neq 0$, the hybridization between the MBSs causes the formation of a fermionic state that couples to both leads. Consequently, $I_{L}$ is suppressed as $\Gamma_R$ diminishes. For larger values of $|\delta E_0|$, this suppression is more pronounced, as the correlation between the left and right sides is strengthened. These results corroborate the findings of Ref.~\cite{dourado2024nonlocality}, where it was found that the suppression of $G_{LL}$ as $\Gamma_R \to 0$ increases when the ratio $|\delta E_0|/T$ is larger. 

We now fix the value of $|\delta E_0| = 0.6 \delta$ and take several values of temperature in Fig.~\ref{energies}(b). As anticipated, when the temperature increases, the correlation caused by the hybridization is less pronounced. We observe that for $T \geq \delta/15$, the effects of the hybridization in the $I_{L}(\Gamma_R)$ have been completely smeared. We note that in Figs.~\ref{SpinfulKitaev2}(d) and \ref{3siteFullmodel}(d), a temperature of $\sim\Delta/62$ is small enough to allow for a MP mapping using local transport. In Fig.~\ref{energies}(c), we show that increasing the voltage bias has a similar effect of reducing the dependence of $I_{L}$ with $\Gamma_R$ for finite $\delta E_0$, compare, for instance, with the results shown in Fig.~\ref{energies}(a). Finally, we explore the dependence of $I_{L} (\Gamma_R)$ with $\Gamma_L$. For $\delta E_0 = 0.6 \delta$, we observe that the value of $I_{L} (\Gamma_R = 0)$ presents a non-monotonic behavior with $\Gamma_L$, as for both small and larger values of $\Gamma_L$, see the red and black lines, the value of $I_{L} (\Gamma_R = 0)$ is larger than the intermediary cases, depicted by the green, blue, and purple curves. 

\begin{figure}
\centering
\includegraphics[width=\linewidth]{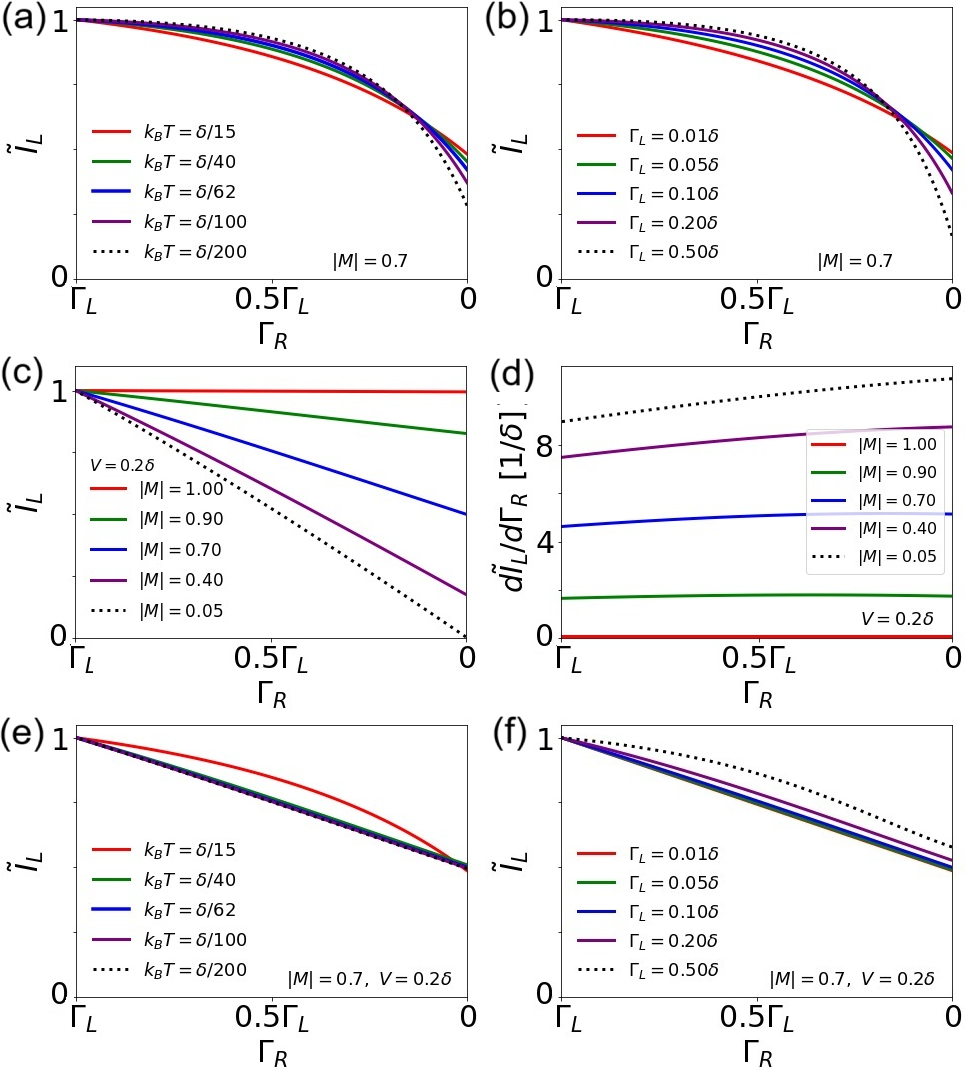}
\caption{(a) $\Tilde{I}_{L}$ and (b) $d \Tilde{I}_{L}/d \Gamma_R$ as functions of $\Gamma_R$ for different values of $T$ and $\Gamma_L$, respectively, at $V_L = -V_R = 0.01\delta$ an $|M| = 0.7$. (c)-(d) $\Tilde{I}_{L}$ and $d \Tilde{I}_L/d\Gamma_R$ for distinct values of $|M|$ at $V_L = -V_R =0.2\delta$. (e)-(f) Same as (a) and (b), respectively, considering larger voltage bias $V_L = -V_R = 0.2\delta$.}
\label{MPsExploringParameters}
\end{figure}

We next study the effect of lowering MP. Differently from the previously discussed case with $\delta E_0\neq0$, the value of $I_{L}(\Gamma_R = 0)$ for $|M| < 1$ saturates at a point, proportional to the value of the MP for larger temperatures. To see this, we fix $\delta E_0 = 0$ and $|M| = 0.7$, blue cross in Fig.~\ref{Kitaev2}(a), and take several values of $T$. We observe in Fig.~\ref{MPsExploringParameters}(a) that lower temperatures increase the suppression of $I_{L}$ at lower $\Gamma_R$, and as the temperature increases, the current tends to flatten out, reaching a value predicted by Eq.~(\ref{measuredMP}). We obtain a similar dependence with $\Gamma_L$ for a fixed temperature: lower values of $\Gamma_L$ cause the $I_{L}(\Gamma_R)$ to flatten out, as shown in Fig.~\ref{MPsExploringParameters}(b). Therefore, low values of $\Gamma_L$ and/or large values of $T$ cause the results to saturate at a value described by Eq.~(\ref{measuredMP}). On the other hand, increasing $\Gamma_L$ and diminishing $T$ significantly reduce the value of $I_{L} (\Gamma_R = 0)$, beneficial to identify parameter regions with high MP. 

For large bias, the results become independent of $\Gamma_L$ and $T$, provided that the regime $\Gamma_L, T \ll V_L \ll E_1 - E_0$, where $E_1 - E_0$ is the excitation gap, is respected. Essentially, the derivative of the Fermi function in Eq.~(\ref{GL}) will sample larger energy values in the transmission and reflection amplitudes. Since $\Gamma_{L, R}$ is the smallest energy scale, the results become linear in $\Gamma_R$ as higher order terms become negligible. The simulations of $I_{L}$ and $d I_{L}/ d \Gamma_R$ as a function of $\Gamma_R$ in Figs.~ \ref{MPsExploringParameters}(c, d) corroborate the linear dependence discussed above. We note in addition that $d I_{L}/d \Gamma_R$ becomes independent of the ratio $\Gamma_R/\Gamma_L$. Finally, the regime of large bias makes the results independent of the values of $\Gamma_R$ and $\Gamma_L$, as shown in Figs.~\ref{MPsExploringParameters}(e, f), respectively. Since the effects of large voltage bias come from increasing the value of the energy in $A_L (E)$, $T_{RL} (E)$, and $A_{RL}$ in Eq.~(\ref{GL}), the results are similar to the ones for the regime $\Gamma_{L, R} \ll T \ll \delta, \Delta$ presented in Fig.~\ref{Kitaev2}(d), solid red and dotted black lines, and Figs.~\ref{SpinfulKitaev2}(c-f).  

\section{Coherence factors and rate equations} \label{AppCoherenceF}

In this section, we associate the current ratio shown in the main text with the coherence factors of the Bogoliubov operators. Initially, we consider that the leads couple to a single spinless ABS, which is compatible with the results for the Kitaev chain and the microscopic model at large Zeeman fields. Then, we expand the analysis to include both spin channels. This case becomes important when the magnetic field is weak ($V_z < \Delta$).

\subsection{Spinless ABS}

Let us first consider a general spinless ABS such that the fermionic operator regarding the left (right) QD can be expressed as $c_{1 (3)} = u_{L (R)} \alpha + v_{L (R)} \alpha^\dagger$, where the excitations are $\ket{1} = \alpha^\dagger \ket{0}$. Considering the following tunneling Hamiltonian

\begin{equation}
    H_T = \sum_k \left(t_L c_1^\dagger f_{k,L} + t_R f_{k, R}^\dagger c_3 + {\rm H.c.} \right),
\end{equation}
where $f_{k, \alpha}$ destroys an electron of momentum $k$ in lead $\alpha = L, R$. The calculation of the current is greatly simplified if we consider the following parameter regime, $V = V_L/2 = -V_R/2 \gg T \gg \Gamma_{L,R}$. The large voltage bias allows us to neglect the processes of creation and destruction of electrons on the left and right leads, respectively. In addition, for small couplings to the leads we can consider only first-order transitions in $\Gamma_{L, R}$.  

To calculate the current, we use the rate equations, where we calculate the probability of the system to be in states $\ket{0}$ and $\ket{1}$. The occupation probability fulfills 
\begin{equation} \label{ProbConserv}
    P_0 + P_1 = 1.
\end{equation}
In the steady state, we have
\begin{equation} \label{dynamicsP0}
    \dot{P}_0 = \omega_{01} P_1 -\omega_{10} P_0 = 0,
\end{equation}
where $\omega_{a a'}$ is the transition rate from state $\ket{a'}$ to $\ket{a}$, given by
\begin{equation}
\begin{split}
    \omega_{0 1} &= 2 \pi \rho |\bra{0} t_L \Tilde{d}_{k, L} \left(u_L^* \alpha^\dagger +v_L^* \alpha \right)\ket{1}|^2  +\\ 
    &+2 \pi \rho |\bra{0}t_R \Tilde{d}_{k, R}^\dagger \left(u_R \alpha + v_R \alpha^\dagger \right)\ket{1}|^2 \,,
\end{split}
\end{equation}
such that
\begin{equation}
    \omega_{0 1} =  \omega_{01}^L + \omega_{0 1}^ R = \Gamma_L |v_L|^2 + \Gamma_R |u_R|^2,
\end{equation}
where $\rho$ is the density of states in the leads, assumed constant in the considered energy range, and $\Gamma_a = 2 \pi \rho |t_a|^2$. We also distinguish between the contributions from left and right sides to $\omega_{0 1}$, which will be important when we calculate the current on one of the sides. Similarly,

\begin{equation}
    \omega_{1 0 } = \Gamma_L |u_L|^2 + \Gamma_R |v_R|^2.
\end{equation}
From Eqs. (\ref{ProbConserv}) and (\ref{dynamicsP0}), we obtain

\begin{equation}
    P_0 = \frac{\omega_{01}}{\omega_{01} + \omega_{10}},
\end{equation}
\begin{equation}
    P_1 = \frac{\omega_{10}}{\omega_{01} + \omega_{10}}.
\end{equation}

Substituting the expressions for $\omega_{01}$ and $\omega_{10}$ into the above equations, we obtain

\begin{equation}
    P_0 = \frac{\Gamma_L |v_L|^2 + \Gamma_R |u_R|^2}{\Gamma_L \left(|u_L|^2 + |v_L|^2 \right) + \Gamma_R \left(|u_R|^2 + |v_R|^2 \right)}
\end{equation}
\begin{equation}
    P_1 = \frac{\Gamma_L |u_L|^2 + \Gamma_R |v_R|^2}{\Gamma_L \left(|u_L|^2 + |v_L|^2 \right) + \Gamma_R \left(|u_R|^2 + |v_R|^2 \right)}.
\end{equation}

Finally, we obtain the current on the left lead by counting the particles tunneling into $d_1$ by multiplying the tunneling rates on each side by the probabilities of occupation for each initial state.

\begin{equation}
    I_L = \omega_{10}^L P_0 + \omega_{0 1}^L P_1,
\end{equation}
\begin{equation} \label{IL_complete}
    I_L = \Gamma_L \frac{2 \Gamma_L |u_L|^2 |v_L|^2 + \Gamma_R \left( |u_L|^2 |u_R|^2 + |v_L|^2 |v_R|^2 \right)}{\Gamma_L \left(|u_L|^2 + |v_L|^2 \right) + \Gamma_R \left(|u_R|^2 + |v_R|^2 \right)}.
\end{equation}
For $\Gamma_R = 0$,

\begin{equation} \label{ILGRZeroApp}
    I_L (\Gamma_R = 0) = \Gamma_L \frac{2 |u_L|^2 |v_L|^2}{|u_L|^2 + |v_L|^2}.
\end{equation}

For $\Gamma_R = \Gamma_L$,

\begin{equation}
    I_L (\Gamma_R = \Gamma_L) = \Gamma_L \frac{2 |u_L|^2 |v_L|^2 + |u_L|^2 |u_R|^2 + |v_L|^2 |v_R|^2}{|u_L|^2 + |v_L|^2 + |u_R|^2 + |v_R|^2}.
\end{equation}
We first consider a configuration with symmetric coherence factors, $|u_L| = |u_R| = |u|$ and $|v_L| = |v_R| = |v|$ (see App.~\ref{AppGLR} for a discussion on the role of asymmetry), obtaining

\begin{equation} \label{ILSpinless}
    \frac{I_L (\Gamma_R = 0)}{I_L (\Gamma_R = \Gamma_L)} = \left(\frac{2 |u| |v|}{|u|^2 + |v|^2} \right)^2.
\end{equation}
For MBSs, where $|u| = |v|$, $\frac{I_L (\Gamma_R = 0)}{I_L (\Gamma_R = \Gamma_L)} = 1$. As the MP diminishes, we will observe a reduction for the product $|u||v|$. For instance, $|u| = 0$ or $|v| = 0$ (purely fermionic state), $\frac{I_L (\Gamma_R = 0)}{I_L (\Gamma_R = \Gamma_L)} = 0$. Therefore, the MP is associated with the coherence factors and can be probed via local conductance or current measurements. In App.~\ref{AppGLR}, we relax the symmetry requirement and discuss results for asymmetric setups.

In general, our method relies on performing two consecutive measurements to estimate the MP, meaning other expressions can be developed to this end. For instance, from Eqs.~(\ref{IL_complete}) and (\ref{MPsigma}), considering also a symmetric profile ($u_L = u_R = u$ and $v_L = v_R = v$) it is possible to show that 

\begin{equation}
    |M|^2 = \frac{I_L (\Gamma_R = 0)}{I_L(\Gamma_R = 0) + I_L(\Gamma_R \gg \Gamma_L)},
\end{equation}
which only relies on measuring the current in the regimes $\Gamma_R \ll \Gamma_L$ and $\Gamma_R \gg \Gamma_L$. Another example is to consider small deviations for the coupling to one of the leads and calculate the derivative. For this case, we obtain

\begin{equation}
    \frac{d I_L}{d\Gamma_R} = \left(\frac{\Gamma_L}{\Gamma_L + \Gamma_R} \right)^2 \left(1 - |M|^2 \right).
\end{equation}
This expression has the advantage of requiring only small deviations on $\Gamma_R$, but it requires knowledge of the values of $\Gamma_{L, R}$.

\subsection{Finite Zeeman - Spinful ABS}

Now we consider the spinful case, which becomes relevant when the Zeeman energy becomes smaller than the gap ($V_z < \Delta$). In this case, the Bogoliubov operator for a generic ABS is given by $c_{1\sigma} = c_{3\sigma} = u_\sigma \alpha_\sigma + v_\sigma \alpha_\sigma^\dagger$. The excitations are given by $\ket{\uparrow} = \alpha_\uparrow^\dagger \ket{0}$, $\ket{\downarrow} = \alpha_\downarrow^\dagger \ket{0}$, and $\ket{\uparrow \downarrow} = \alpha_\uparrow^\dagger \alpha_\downarrow^\dagger \ket{0}$.

We calculate the transition rates as before, $\omega_{ij}^r = 2\pi \rho |\bra{i} H_T^r \ket{j}|^2$, $r = L, R$, obtaining

\begin{equation} \label{omegasBeg}
    \omega_{0 \uparrow} = \Gamma_L |v_\uparrow|^2 + \Gamma_R |u_\uparrow|^2,
\end{equation}
\begin{equation}
    \omega_{\uparrow 0} = \Gamma_L |u_\uparrow|^2 + \Gamma_R |v_\uparrow|^2,
\end{equation}
\begin{equation}
    \omega_{0 \downarrow} = \Gamma_L |v_\downarrow|^2 + \Gamma_R |u_\downarrow|^2,
\end{equation}
\begin{equation} \label{omegasEnd}
    \omega_{\downarrow 0} = \Gamma_L |u_\downarrow|^2 + \Gamma_R |v_\downarrow|^2,
\end{equation}
where we consider left-right symmetric coefficients. In addition, we verify that $\omega_{2 \uparrow} = \omega_{\downarrow 0}$, $\omega_{\uparrow 2} = \omega_{0 \downarrow }$, $\omega_{2 \downarrow} = \omega_{\uparrow 0}$, and $\omega_{\downarrow 2} = \omega_{0 \uparrow}$.

The conservation of probability gives us the below relation.

\begin{equation}
    P_0 + P_\uparrow + P_\downarrow + P_{\uparrow \downarrow} = 1.
\end{equation}

In the steady state, we have the additional relations

\begin{equation}
    \dot{P}_0 = \omega_{0\uparrow} P_\uparrow + \omega_{0\downarrow} P_\downarrow - (\omega_{\uparrow 0} + \omega_{\downarrow 0}) P_0 = 0,
\end{equation}
\begin{equation}
    \dot{P}_\uparrow = \omega_{\uparrow 0 } P_0 + \omega_{0\downarrow} P_{\uparrow \downarrow} - (\omega_{ 0 \uparrow } + \omega_{\downarrow 0}) P_\uparrow = 0,
\end{equation}
\begin{equation}
    \dot{P}_{\uparrow \downarrow} = \omega_{\uparrow 0 } P_\uparrow + \omega_{\uparrow 0 } P_\downarrow - (\omega_{ 0 \uparrow } + \omega_{0 \downarrow }) P_{\uparrow \downarrow} = 0,
\end{equation}

where we have considered the large bias voltage limit (larger than the energy of the state and the charging energy in the system). By solving the above system of linear equations, we obtain

\begin{equation} \label{PsBeg}
    P_0 = \frac{\omega_{0 \uparrow} \omega_{0 \downarrow}}{\Omega},
\end{equation}
\begin{equation}
    P_\uparrow = \frac{\omega_{ \uparrow 0 } \omega_{0 \downarrow}}{\Omega},
\end{equation}
\begin{equation}
    P_\downarrow = \frac{\omega_{0 \uparrow} \omega_{ \downarrow 0}}{\Omega},
\end{equation}
\begin{equation} \label{PsEnd}
    P_{\uparrow \downarrow} = \frac{\omega_{ \uparrow 0 } \omega_{ \downarrow 0 }}{\Omega},
\end{equation}
where $\Omega = (\omega_{0 \uparrow  } + \omega_{ \uparrow 0 })(\omega_{ 0 \downarrow } + \omega_{ \downarrow 0 }) = (\Gamma_L + \Gamma_R)\left(|u_\uparrow|^2 + |v_\uparrow|^2 \right)\left(|u_\downarrow|^2 + |v_\downarrow|^2 \right)$.

Now we calculate the current by counting the particles tunneling into $c_1$.

\begin{equation}
\begin{split}
    I_L =& \left(\omega_{\uparrow 0}^L + \omega_{\downarrow 0}^L \right) P_0 + \left(\omega_{0 \uparrow }^L + \omega_{ \downarrow 0}^L \right) P_\uparrow + \left(\omega_{ 0 \downarrow }^L + \omega_{\uparrow 0}^L \right) P_\downarrow +\\
    &+ \left(\omega_{0 \uparrow }^L + \omega_{0 \downarrow }^L  \right) P_{\uparrow \downarrow}.
\end{split}
\end{equation}
Substituting Eqs. (\ref{omegasBeg}-\ref{omegasEnd}) and (\ref{PsBeg}-\ref{PsEnd}) into the above equation, we obtain

\begin{equation}
    I_L = \frac{\Gamma_L}{\Gamma_L + \Gamma_R} \sum_\sigma \frac{2 |u_\sigma|^2 |v_\sigma|^2 \Gamma_L + \left( |u_\sigma|^4 +  |v_\sigma|^4\right) \Gamma_R}{|u_\sigma|^2 +  |v_\sigma|^2}.
\end{equation}
Now we compare the values of the current with the right lead pinched off and having the same coupling as the left lead, i.e.,

\begin{equation}
    I_L (\Gamma_R = 0) = 2 \Gamma_L \sum_\sigma \frac{|u_\sigma|^2 |v_\sigma|^2}{|u_\sigma|^2 + |v_\sigma|^2},
\end{equation}
and
\begin{equation}
    I_L (\Gamma_R = \Gamma_L) = \frac{\Gamma_L}{2}\left(|u_\uparrow|^2 + |v_\uparrow|^2 + |u_\downarrow|^2 + |v_\downarrow|^2 \right),
\end{equation}
Therefore, 

\begin{equation} \label{ILSpinful}
    \frac{I_L (\Gamma_R = 0)}{I_L(\Gamma_R = \Gamma_L)} = \frac{4}{\sum_\sigma |u_\sigma|^2 + |v_\sigma|^2} \sum_\sigma \frac{|u_\sigma|^2 |v_\sigma|^2}{|u_\sigma|^2 + |v_\sigma|^2}.
\end{equation}

As one increases the magnetic field, the spin-up components of the ABS become smaller and, eventually, the system reaches the limit where $|u_\uparrow|, |v_\uparrow| \ll |u_\downarrow|, |v_\downarrow|$. At this point, the system is polarized, with Eq. (\ref{ILSpinful}) becoming approximately Eq. (\ref{ILSpinless}) for the spinless model. 

To visualize how the magnetic field diminishes the spin-up component of the wave function at low energies, let us consider a simple model consisting of a single QD under the influence of a local pairing amplitude and a Zeeman field. For simplicity, we set the QD level to zero and consider all energies in unities of $\Delta$. The Hamiltonian reads

\begin{equation}
    H = V_z c_\uparrow^\dagger c_\uparrow + \Delta \left(c_\uparrow^\dagger c_\downarrow^\dagger + {\rm H.c.} \right).
\end{equation}
In the single particle picture, the lowest-energy sector is characterized by the energies $E_\pm = \pm \frac{1}{2}\left(\sqrt{V_z^2 + 4} - V_z \right)$. The associated eigenvectors are (up to a normalization factor) $\psi_+ = \begin{pmatrix} |E_\pm| & 0 & 0 & 1 \end{pmatrix}$ and $\psi_- = \begin{pmatrix} 0 & |E_\pm| & 1 & 0 \end{pmatrix}$, where we used the basis $\{c_\uparrow, c_\uparrow^\dagger, c_\downarrow, c_\downarrow^\dagger \}$. As $V_z$ increases, the energies and spin-up components of the wave functions asymptotically approach zero, as the system becomes polarized.

\section{Asymmetric setups and Nonlocal Conductance} \label{AppGLR}

\begin{figure}
\centering
\includegraphics[width=\linewidth]{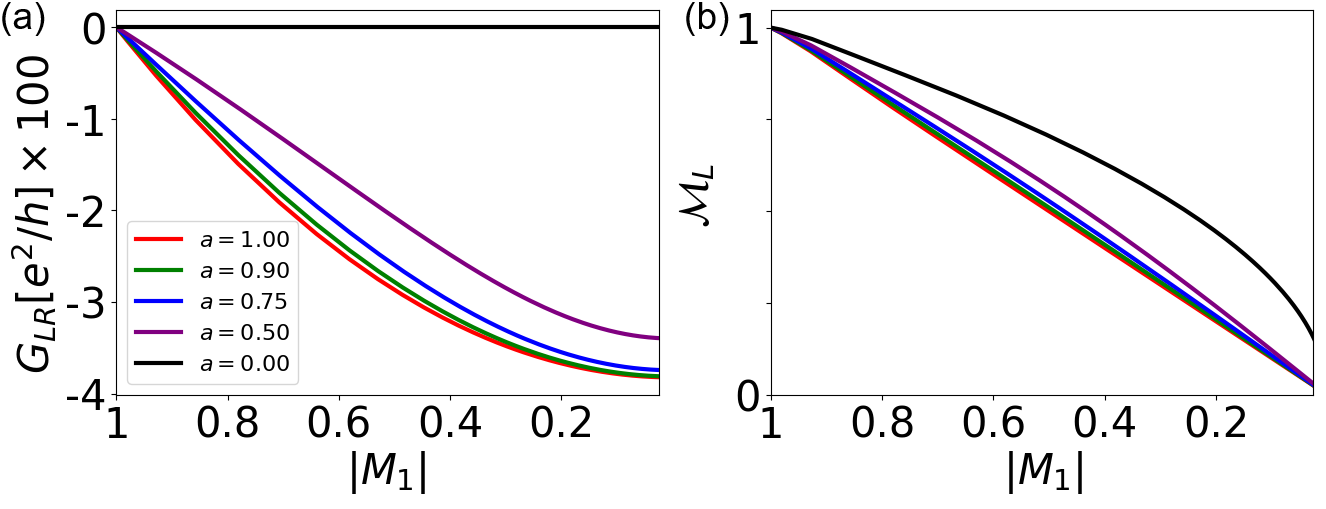}
\caption{(a) Nonlocal conductance $G_{LR}$ as a function of $|M_1|$ and $|M_2|$. The parameter $a$ quantifies the asymmetry in the MPs, when $a = 1$, $|M_1|$ and $|M_2|$ decay equally. For $a < 1$, $|M_2|$ is suppressed at a slower rate, as $\mu_1 = a \mu_2$. For $a = 0$, $|M_2|$ remains constant while $|M_1|$ diminishes. (b) $\mathcal{M}_L$ as a function of $|M_1|$. When considering asymmetric configurations, the results for $\mathcal{M}_L$ are similar, showing that this quantity is essentially a local probe for the MP.}
\label{GLRGLLcomparison}
\end{figure}

In the main text, we discussed how the measurement of the MP is essentially probing the coherence factors. Moreover, in Eq.~(\ref{ILGRZero}) we showed that $I_L (\Gamma_R = 0)$ only depends on the left side of the system, $u_L, v_L$. Here we consider asymmetries in our setup, $\mu_1 \neq \mu_2$, and their impact on $\mathcal{M}$. Moreover, we consider the nonlocal conductance as a possible probe for the MP, verifying that $G_{LR}$ probes the MP at both ends of the Kitaev chain simultaneously.

When considering different chemical potentials for the QDs, the degeneracy line is now given by $t = \sqrt{1 + \mu_1 \mu_2}$. We will parameterize the chemical potentials in the following way, $\mu_1 = a \mu_2 = a \mu$. In this case, the MPs are \cite{Tsintzis2024}

\begin{equation}
    |M_1| = \frac{2\sqrt{1 + a \mu^2}}{2 + (1 + a) \mu^2},
\end{equation}

\begin{equation}
    |M_2| = \frac{2 \sqrt{1 + a \mu^2}}{2 + a(1 + a)\mu^2}.
\end{equation}

Now we choose several values of $a$ and increase $\mu$ to diminish $|M_1|$ and also $|M_2|$ at a different rate depending on the value of $a$. In particular, for $a = 1$ the setup becomes symmetric and we regain the results discussed in the main text. For $a = 0$, $\mu_1 = 0$ while $\mu_2$ increases, which keeps $|M_2| = 1$ while $|M_1|$ decreases. In Fig.~\ref{GLRGLLcomparison}(a), we plot the nonlocal conductance as a function of $|M_1|$. We observe that for the symmetrical setup, $a = 1$, depicted by the red line, the nonlocal conductance vanishes at the sweet spot ($|M| =  1$) and achieves a finite (negative) value as we reduce $|M_1| = |M_2|$. This effect arises because MBSs provide a direct channel for incoming electrons to tunnel across the system. As we take smaller values of $a$, the amplitude of $G_{LR}$ when the MP decreases becomes smaller, until $a = 0$, where $G_{LR} = 0$ for any value of $|M_1|$. We conclude, therefore, that $G_{LR}$ is affected by both $|M_1|$ and $|M_2|$ at the same time. If either $|M_1| = 1$ or $|M_2| = 1$, the nonlocal conductance is unable to detect variations on the MP at the opposite side. This result resonates with the findings of Ref. \cite{danon2020nonlocal}, where it was found that the nonlocal conductance is proportional to the product of the BCS charges $q_L q_R$, where $q_\alpha = |u|^2_\alpha - |v|^2_\alpha$. Therefore, if the $|M_\alpha| = 1$, the BCS charge vanishes on that end of the Kitaev chain, causing the nonlocal conductance to vanish.

The asymmetries in the chemical potentials do not significantly change the MP estimate $\mathcal{M}$ in comparison to the results for $G_{LR}$ . We corroborate this by plotting $\mathcal{M}$ as a function of $|M_1|$ for several values of $a$ in Fig.~\ref{GLRGLLcomparison}(b). We note that most of the lines coincide for the majority of the MP values, apart from minor deviations for $|M_1| < 0.5$. This result was anticipated when we calculated the current at Eq.~(\ref{ILGRZero}), which was solely dictated by the coherence factors, thus the MP, on the left side. In addition, this result is consistent with the findings of Ref.~\cite{danon2020nonlocal}, where the local conductance is shown to depend on the local BCS charge, $q_\alpha$.

In contrast to the nonlocal conductance, $\mathcal{M}_L$ is close to $|M_1|$ in asymmetrical cases, representing a viable measurement. However, for large asymmetries ($a < 0.5$), $\mathcal{M}$ overestimates $|M|$, similar to the results for weaker magnetic fields, see Sec.~\ref{weakVz}. In these cases, our measurement can still be used to maximize the MP. In conclusion, while Eq.~(\ref{measuredMP}) provides an approximately local measurement of the MP, by considering only local coherence factors, the nonlocal conductance probes the MP at both ends of the Kitaev chain.

\bibliography{bibliography.bib}

\end{document}